\documentclass[aps,prd,nofootinbib,twocolumn,letterpaper,preprintnumbers]{revtex4}
\pdfoutput=1
\usepackage{amssymb,amsmath,hyperref,slashed,color,graphicx,bm}
\usepackage[T1]{fontenc}

\hypersetup{colorlinks,citecolor= nicegreen,linkcolor= nicered}
\definecolor{nicered}{rgb}{0.7,0.1,0.1}
\definecolor{nicegreen}{rgb}{0.1,0.5,0.1}

\newcommand{\be}{\begin{equation}}
\newcommand{\ee}{\end{equation}}
\newcommand{\bea}{\begin{eqnarray}}
\newcommand{\eea}{\end{eqnarray}}
\newcommand{\Eq}[1]{Eq.~(\ref{#1})}

\newcommand{\Sec}[1]{Sec.~(\ref{#1})}

\newcommand{\Fig}[1]{Fig.~(\ref{#1})}

\begin{document}
\pagestyle{plain}

\title{Electroweak Cogenesis}

\author{Clifford Cheung}
\affiliation{California Institute of Technology, Pasadena, CA 91125}
\author{Yue Zhang}
\affiliation{California Institute of Technology, Pasadena, CA 91125}

\begin{abstract}
We propose a simple renormalizable model of baryogenesis and asymmetric dark matter generation at the electroweak phase transition.   Our setup utilizes the two Higgs doublet model plus two complex gauge singlets, the lighter of which is stable dark matter.  The dark matter is charged under a global symmetry that is broken in the early universe but restored during the electroweak phase transition.
Because the ratio of baryon and dark matter asymmetries is controlled by model parameters, the dark matter need not be light.  Thus,  new force carriers are unnecessary and the symmetric dark matter abundance can be eliminated via Higgs portal interactions alone.   Our model places a rough upper bound on the dark matter mass, and has implications for direct detection experiments and particle colliders. 
\end{abstract}

\preprint{CALT-68-2940}

\maketitle 

\section{Introduction}

Asymmetric dark matter (ADM) is an elegant framework that postulates a common origin for the baryon asymmetry of the universe (BAU) and relic dark matter (DM).  In ADM~\cite{Nussinov:1985xr, reviews,marylandADM}, the observed abundance of DM carries an imbalance between particles and anti-particles seeded by a dynamical link between DM number, $U(1)_{\rm DM}$, and baryon number, $U(1)_{\rm B}$, during an earlier cosmological epoch. 

In the very simplest models of ADM, the baryon and DM asymmetries are equal up to rational numerical coefficients arising from the constraints of chemical equilibrium~\cite{chemicalpot}, implying that $m_{\rm DM}\sim \textrm{GeV}$.  For such low masses, new light mediators are required to efficiently annihilate away the symmetric component of the DM.  However, these theories offer few clues to the so-called coincidence problem, which asks why $\Omega_{\rm DM}/\Omega_{\rm B} \sim 5$.   In particular, there is no reason that $m_{\rm DM}$ should be so tantalizingly close to the mass of the proton.

On the other hand, it is known that more elaborate theories of ADM can accommodate $m_{\rm DM} \gg \textrm{GeV}$ when the mechanism of cogenesis entails model parameters that separately control the asymmetries in baryons and DM~\cite{Buckley:2010ui,Falkowski:2011xh}.  This suggests an underlying reason for the coincidence problem: the DM mass is intrinsically connected to the weak scale and {\it not} the GeV scale.  In turn, the baryon and DM asymmetries are accommodated within the numerical slop of the model parameter space.  This ADM ``miracle'' parallels the so-called weakly interacting massive particle (WIMP) ``miracle'', which famously exploits a coincidence in the relative values of the weak and Planck scales.

In this paper we propose a simple model of ``electroweak cogenesis'' which simultaneously generates the BAU and ADM during the electroweak phase transition (EWPT).  DM is inextricably tied to the weak scale through its participation in the EWPT.  Because the DM is a weak scale particle, light mediator particles are unnecessary to eliminate the symmetric abundance of DM.  Instead, we pursue a considerably simpler setup in which DM annihilates exclusively through Higgs portal interactions.
As a result, electroweak cogenesis is quite minimal and can be achieved in the two Higgs doublet model (2HDM) augmented by two complex scalars charged under an exact DM symmetry, $U(1)_{\rm DM}$.

A novel aspect of our scenario is the pattern of symmetry breaking and restoration during the EWPT:
\bea
SU(2)_{\rm L} \times U(1)_{\rm Y} \rightarrow  U(1)_{\rm EM} \times U(1)_{\rm B}\times U(1)_{\rm DM}.
\label{eq:symmetry}
\eea
At high temperatures, the Higgs doublets take on vanishing vacuum expectation values (VEVs) and electroweak symmetry is preserved.  Electroweak sphalerons are in equilibrium and $U(1)_{\rm B}$ is violated.  Meanwhile, the complex singlets are initialized with non-zero VEVs, so $U(1)_{\rm DM}$ is spontaneously broken in the early universe.  However, at low temperatures, the Higgs doublets break electroweak symmetry and the complex singlets return to the origin of field space.  Hence, $U(1)_{\rm B}$ and $U(1)_{\rm DM}$ are simultaneously {\it restored} during the EWPT.  

Unlike the majority of existing ADM models, which entail high scale dynamics or non-renormalizable operators, our setup achieves cogenesis with renormalizable interactions among weak scale particles.  As a consequence, this model has direct implications for experimental probes in direct detection and high energy colliders.

The outline of this paper is as follows.  In \Sec{sec:model}, we define the particles and interactions for the minimal model of electroweak cogenesis.  Afterwards, we detail our cosmological scenario in \Sec{sec:cosmo}, focusing on the dynamics of the EWPT and the asymmetries generated in baryons and ADM.   Finally, in \Sec{sec:exp}, we discuss experimental constraints from direct detection, and conclude in \Sec{sec:conclusions}.

\section{Model}\label{sec:model}

What is the simplest model of cogenesis?  In this section we address this question systematically.  As is well-known, ADM requires two basic ingredients: {\it i)} a symmetry to protect the asymmetric abundance of DM, and {\it ii)} a strong annihilation channel to deplete the symmetric abundance of DM.  

Ingredient {\it i)} requires that DM be a complex scalar or Dirac fermion charged under $U(1)_{\rm DM}$.   Is it possible to identify $U(1)_{\rm DM}$ with an existing symmetry of the SM, {\it e.g.}~baryon or lepton number?  The only way to imbue a neutral DM particle with baryon or lepton number is to couple it to the SM via higher dimension operators.  While higher dimension operators are often employed in models of ADM and even certain theories of baryogenesis \cite{Cheung:2013hza}, we forgo them here in search of a fully renormalizable model of ADM.  Higgs number is another logical possibility, but  electroweak symmetry breaking will induce late time  DM oscillations~\cite{Buckley:2011ye} which will instantaneously erase the DM asymmetry unless the associated Higgs couplings are exceedingly weak or non-renormalizable~\cite{Blum:2012nf,Servant:2013uwa}.  For these reasons, we take $U(1)_{\rm DM}$ to be an additional global symmetry beyond the SM.
Note that DM asymmetry generation requires that $U(1)_{\rm DM}$ be broken in the early universe, so to avoid the washout of the DM asymmetry, $U(1)_{\rm DM}$ must be restored fast enough at late times.

Ingredient {\it ii)} is typically achieved via DM interactions with light mediator particles.  While such theories offer rich possibilities for ``dark sector'' model building and phenomenology~\cite{ArkaniHamed:2008qn}, they also introduce considerable complexity, {\it e.g.}~dark forces, dark Higgs bosons, and dark cosmology.  In the present work, we eschew light mediators in the interest of minimality, and insist that DM annihilates into SM particles directly. If DM annihilates via gauge interactions, stringent limits from direct detection~\cite{Cirelli:2005uq} have already excluded complex scalar/Dirac fermion DM with non-zero hypercharge.  On the other hand, DM with zero hypercharge is still allowed, making the complex triplet the lowest viable DM representation. In order to avoid exotic gauge representations, we consider the simpler possibility that DM is a complex gauge singlet scalar interacting via the Higgs portal.

\begin{figure}[t]
\centerline{\includegraphics[width=1.0\columnwidth]{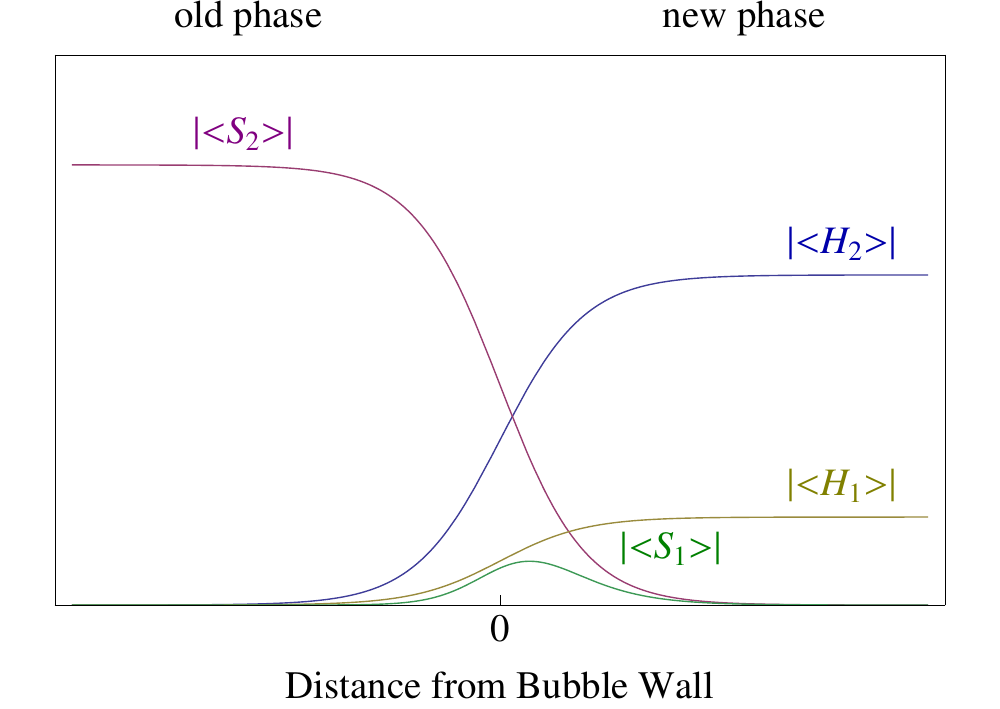}}
\caption{Depiction of VEV profiles across the bubble wall.}\label{pic}
\end{figure}

Our setup---electroweak cogenesis---is a generalization of electroweak baryogenesis that incorporates ADM.   To satisfy the Sakharov conditions, we require violation of $U(1)_B$, $U(1)_{\rm DM}$, and CP invariance during an out-of-equilibrium phase of the early universe.
While the SM offers inherent baryon number violation by way of electroweak sphalerons, it lacks the CP violation and strong first-order EWPT required of electroweak baryogenesis.  For this reason, the very simplest models of electroweak baryogenesis employ the 2HDM \cite{McLerran:1990zh, Turok:1990zg, Fromme:2006cm,Gunion:1989we}, which we take to be our starting point.

A novel aspect of our framework is that $U(1)_{\rm DM}$ is spontaneously broken in the early universe by the VEV of a $U(1)_{\rm DM}$ charge scalar field.  If electroweak sphalerons are to be operative before the EWPT, then this $U(1)_{\rm DM}$ breaking field must be a gauge singlet. 
This suggests a minimal setup in which the $U(1)_{\rm DM}$ breaking VEV is that of a single DM particle beyond the SM.
Unfortunately, this simple possibility cannot work: in a theory with only renormalizable interactions, the existence of an exact, unbroken $U(1)_{\rm DM}$ symmetry linked to a physical CP phase requires at least two new particles charged under the $U(1)_{\rm DM}$.

\smallskip
The minimal incarnation of ADM then emerges---it consists of two Higgs doublets, $H_{1}$ and $H_2$, and two complex singlets, $S_{1}$ and $S_2$, coupled via all gauge invariant, renormalizable interactions consistent with certain global symmetries.  We assume a softly broken parity, $H_1 \rightarrow -H_1$, $H_2\to H_2$, in order to evade stringent constraints from flavor changing neutral currents (FCNCs); this is the conventional choice that defines the Type-II 2HDM.  Moreover, we impose an exact $U(1)_{\rm DM}$ symmetry under which $S_1$ and $S_2$ carry unit charge.  The scalar potential takes the form
\bea
V &=&   V_{H}+V_{S}+V_{H\textrm{-}S},
\label{eq:action0}
\eea
where
$V_H$ describes the masses and self interactions of the Higgs doublets, 
\bea
\label{eq:action}
V_H &=&  m_{1,H}^2 |H_1|^2 + m_{2,H}^2 |H_2|^2  + [m_{3,H}^2  H^\dagger_1 H_2 +\textrm{c.c}] \nonumber\\
&& +\frac{\lambda_{1,H}}{2}  |H_1|^4 + \frac{\lambda_{2,H} }{2} |H_2|^4 +  \lambda_{3,H} |H_1|^2 |H_2|^2 \nonumber \\
&&+ \lambda_{4,H} |H_1^\dagger  H_2|^2 +[ \lambda_{5,H} (H_1^\dagger  H_2)^2 + \textrm{c.c.} ].
\label{eq:VH}
\eea 
Meanwhile, $V_S$ describes the masses and self interactions of the complex singlets, assuming an additional softly broken parity, $S_1 \rightarrow -S_1$, $S_2\to S_2$, so
\bea
V_S &=& m_{1,S}^2 |S_1|^2 + m_{2,S}^2 |S_2|^2  + [m_{3,S}^2  S^\dagger_1 S_2 +\textrm{c.c}] \nonumber\\ 
&& + \frac{\lambda_{1,S}}{2}  |S_1|^4 + \frac{\lambda_{2,S}}{2}  |S_2|^4 +  \lambda_{3,S} |S_1|^2 |S_2|^2 \nonumber \\
&&+[ \lambda_{4,S} (S_1^\dagger  S_2)^2 + \textrm{c.c.} ] ,
\label{eq:VS}
\eea
expressed in a nomenclature for couplings and masses that mirrors that of the usual 2HDM \cite{Fromme:2006cm,Gunion:1989we}.   Finally, $V_{H\textrm{-}S}$ describes the portal connecting the Higgs doublets to the complex scalars,
\bea \label{HSpot}
V_{H\textrm{-}S} &=& \kappa_{1} |H_1|^2|S_1|^2 +\kappa_{2} |H_2|^2|S_2|^2 \nonumber \\
&& +\kappa_{3} |H_1|^2|S_2|^2 + \kappa_{4} |H_2|^2|S_1|^2 \nonumber\\
&& + [ \epsilon_1 H_1^\dagger H_2 S_1^\dagger S_2 + \epsilon_2 H_1^\dagger H_2 S_2^\dagger S_1 + \textrm{c.c.}]. 
\label{eq:VHS}
\eea 
While $U(1)_{\rm DM}$ is exact at the level of the Lagrangian, it will be spontaneously broken in the early universe.

After electroweak symmetry breaking, $V_{H\textrm{-}S}$ induces additional contributions to the masses of $S_1$ and $S_2$.  As a result, $S_1$ and $S_2$ mix by an amount proportional to $\epsilon_1$ and $\epsilon_2$.  However, as we will see later, $\epsilon_1$ and $\epsilon_2$ must be small to accommodate the observed asymmetric abundances of baryons and DM---so $S_1$ and $S_2$ will be approximate mass eigenstates.  We denote the physical masses of these eigenstates by $m_1$ and $m_2$, where without loss of generality we take $m_1 < m_2$.  As a consequence of the unbroken $U(1)_{\rm DM}$ at low energies, $S_2$ decays to $S_1$, which comprises the stable DM.

\section{Cosmology}\label{sec:cosmo}

\subsection{Electroweak Phase Transition}

\label{sec:EWPT}

Electroweak cogenesis requires a strongly first-order EWPT, together with the symmetry breaking pattern of \Eq{eq:symmetry}. 
We assume that the scalar VEVs transit according to
\begin{equation}
\begin{tabular}{c|c|c|c|c}
& $\langle H_1 \rangle$ & $\langle H_2 \rangle$  & $\langle S_1 \rangle$ & $\langle S_2 \rangle$ \\ \hline 
$T \gtrsim T_c$  & $  0$ & $  0 $ & 0  & $w_c$\\
$T \lesssim T_c$  & $0$ & $v_c $ & 0  &0
\end{tabular} 
\label{eq:VEVs}
\end{equation}
where $v_c$ and $w_c$ are the critical VEVs at the critical temperature of the electroweak phase transition, $T_c$.  
For simplicity, we assume that $H_2$ and $S_2$ drive the dominant VEVs during the EWPT.  Indeed, while these fields will in general induce subdominant VEVs for $H_1$ and $S_1$ proportional to $m_{3,H}^2$ and $m_{3,S}^2$, we assume that these parameters are small.  
Concretely, \Eq{eq:VEVs} applies at the leading, zeroth order in powers of $m_{3,H}^2/T_c^2$ and $m_{3,S}^2/T_c^2$; the former is equivalent to large $\tan\beta \equiv \langle H_2 \rangle/\langle H_1 \rangle$ limit.
Since CP violating interactions only enter at next-to-leading order in this expansion, we can neglect their effects on the strength of the first-order phase transition.  Of course, these CP violating effects will be critical for the generation of particle asymmetries, as we will see later.

For our analysis of the phase transition we compute the finite temperature effective potential for $H_2$ and $S_2$,   
\bea
V(T) &=& V(0) + \Delta V(T).
\eea
The zero temperature potential is given by
\bea \label{zeropot}
V(0) &=& m^2_{2,H} |H_2|^2+ \frac{\lambda_{2,H}}{2} |H_2|^4  +m^2_{2,S} |S_2|^2  + \frac{\lambda_{2,S}}{2} |S_2|^4 \nonumber \\
&&+  \kappa_2 |H_2|^2 |S_2^2| + V_{\rm CW}(H_2,S_2),
\eea
where $V_{\rm CW}$ is the one-loop Coleman-Weinberg effective potential computed in the prescription of~\cite{Cline:1996mga,Quiros:1999jp}.

For the thermal potential, we include daisy-resummed thermal cubic contributions and leading logarithmic thermal corrections~\cite{Cline:1996mga,Anderson:1991zb} in the high-T expansion,
\begin{eqnarray}\label{thermalpot}
\Delta V(T) &=& \Delta m^2_{2,H}(T) |H_2|^2+  \Delta m^2_{2,S}(T) |S_2|^2 \\
&&\hspace{-1.5cm} - \frac{T}{12 \pi} {\rm Tr} \, {\cal M}_B (H_2, S_2, T)^3 \nonumber \\ 
&& \hspace{-1.5cm}-   \frac{1}{64\pi^2} {\rm Tr} \,{\cal M}_B^4 (H_2,S_2,0) \left[\log \frac{{\cal M}_B^2(H_2,S_2,0)}{T^2}-c_B\right] \nonumber\\
&& \hspace{-1.5cm}+   \frac{1}{64\pi^2} {\rm Tr}\, {\cal M}_F^4 (H_2,S_2 ,0) \left[\log \frac{{\cal M}_F^2(H_2,S_2,0)}{T^2}-c_F\right], \nonumber
\end{eqnarray}
where ${\cal M}_{B,F}(H_2,S_2,T)$ is the field dependent, thermal mass matrix for all bosons and fermions, respectively, and $c_{B,F}\simeq5.4, 2.6$.  The thermal mass corrections for $H_2$ and $S_2$ are
\begin{eqnarray}\label{thermalmass}
\Delta m_{2,H}^2(T) \!\!&=&\!\! 
\frac{T^2}{12} \left[   ({9g^2+3g'^2})/{4}  + 3y_t^2 + 3 \lambda_{2,H}\rule{0mm}{4mm}\right. \nonumber \\
&& \hspace{.7cm} \left.+ 2\lambda_{3,H}+ \lambda_{4,H}+ \kappa_{2} + \kappa_{4} \rule{0mm}{4mm}\right]   \nonumber \\
\Delta m_{2,S}^2(T) \!\!&=&\!\!  
\frac{T^2}{12} \left[ 2 \lambda_{2,S} + \lambda_{3,S}+ 2 \kappa_{2} + 2 \kappa_{3} \rule{0mm}{4mm}\right] .
\end{eqnarray} 
To compute the thermal potential, we have followed the conventional approach of \cite{Arnold:1992rz}, where the daisy resummation is applied solely to the cubic interaction terms, {\it i.e.}~the second line of \Eq{thermalpot}.  Alternative methods, such as those of \cite{Parwani:1991gq} yield similar numerical results.

\begin{figure}[t]
\centerline{\includegraphics[width=1.0\columnwidth]{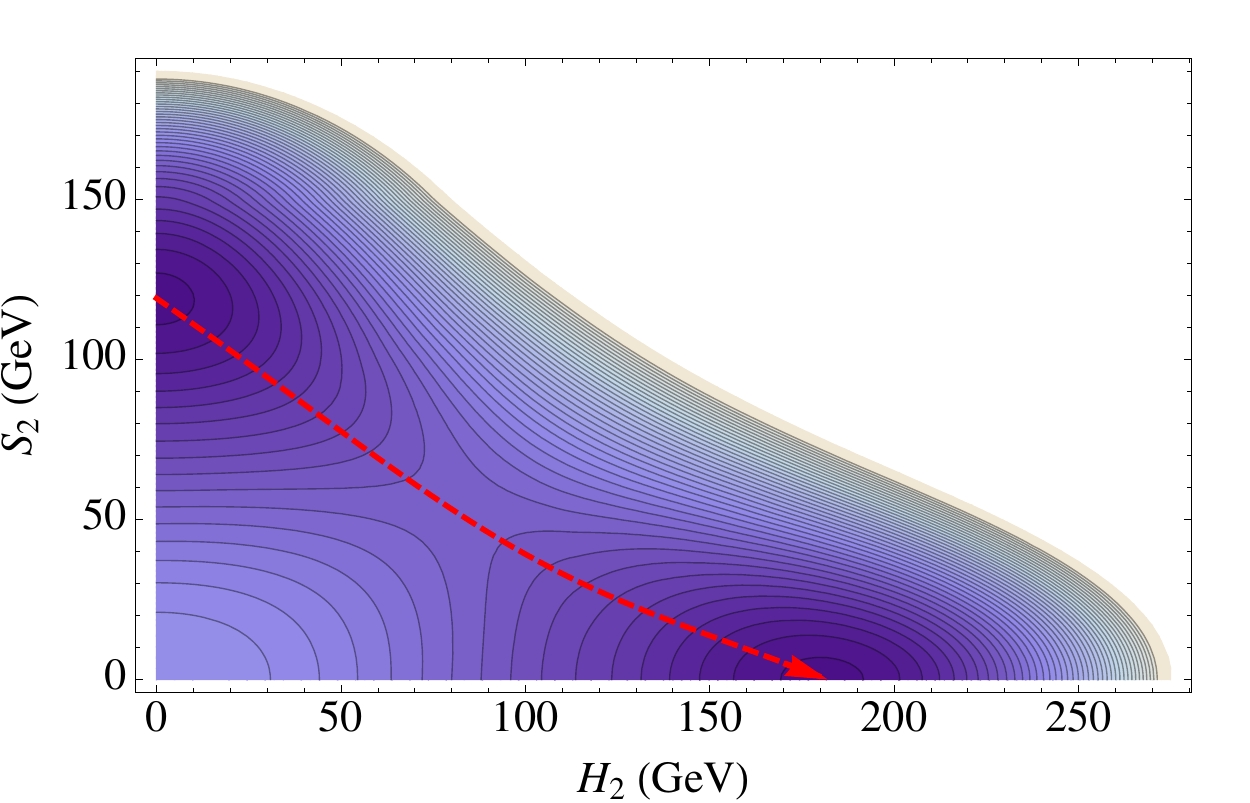}}
\caption{An example scalar potential which yields a successful first-order EWPT. The red dotted line shows the path of the transition, which traverses from the minimum at $\langle S_2\rangle \neq 0$ to the minimum at $\langle H_2 \rangle \neq 0$.}\label{PhT1}
\end{figure} 

Note that despite their common usage, $V_{\rm CW}$ and the thermal cubic and logarithmic terms in \Eq{thermalpot} are gauge non-invariant quantities~\cite{Jackiw:1974cv, Dolan:1973qd, Patel:2011th}, which vary under choice of $R_\xi$ gauges.  Because of these gauge ambiguities, we will present our results with and without the higher order corrections---concretely, for the latter we include only the tree level zero temperature potential plus the gauge invariant thermal masses in \Eq{thermalmass}.  Gauge ambiguities in the thermal cubic and logarithmic potential terms will not qualitatively affect our conclusions, which are driven primarily by the tree level action.  As discussed in \cite{Patel:2011th}, there exists a fully gauge invariant methodology for computing the dynamics of the first-order phase transition, however we leave this analysis to future work.

Our setup requires a strongly first-order phase transition from the $\langle S_2 \rangle \neq0$ vacuum to the $\langle H_2 \rangle \neq 0$ vacuum.  At temperatures just above $T_c$, the $\langle S_2 \rangle \neq 0$ vacuum should be deeper than the $\langle H_2 \rangle \neq 0$ vacuum, and vice versa at temperatures just below $T_c$.  
Thus, $T_c$ is defined as the temperature at which the two vacua in \Eq{eq:VEVs} are degenerate, so
\begin{eqnarray}\label{PTconditions}
V(T_c)\bigg|_{{H_2= 0} \atop {S_2=w_c}} &=& V( T_c)\bigg|_{{H_2= v_c} \atop {S_2=0}} ,
\end{eqnarray}
where a strong first-order EWPT requires that $v_c / T_c >0.9$ for sufficient suppression of the sphaleron rate inside the broken phase.

Note that a large and positive tree-level quartic term $\kappa_2|H_2|^2|S_2|^2$ is crucial for supporting a potential barrier between the two degenerate vacua.  Thus, the requirement of a first-order phase transition places a lower bound on $\kappa_2$. 
This contrasts with the usual picture of the EWPT in which the potential barrier relies on the daisy-resummed thermal cubic term.
As a consequence, the existence of a first-order phase transition does not impose a stringent lower bound~\cite{Fromme:2006cm,Huber2013,Chowdhury:2011ga} on the mass scale of the second doublet.  An example potential which can accommodate the required EWPT is shown in \Fig{PhT1}.

\begin{figure}[t]
\includegraphics[width=1.0\columnwidth]{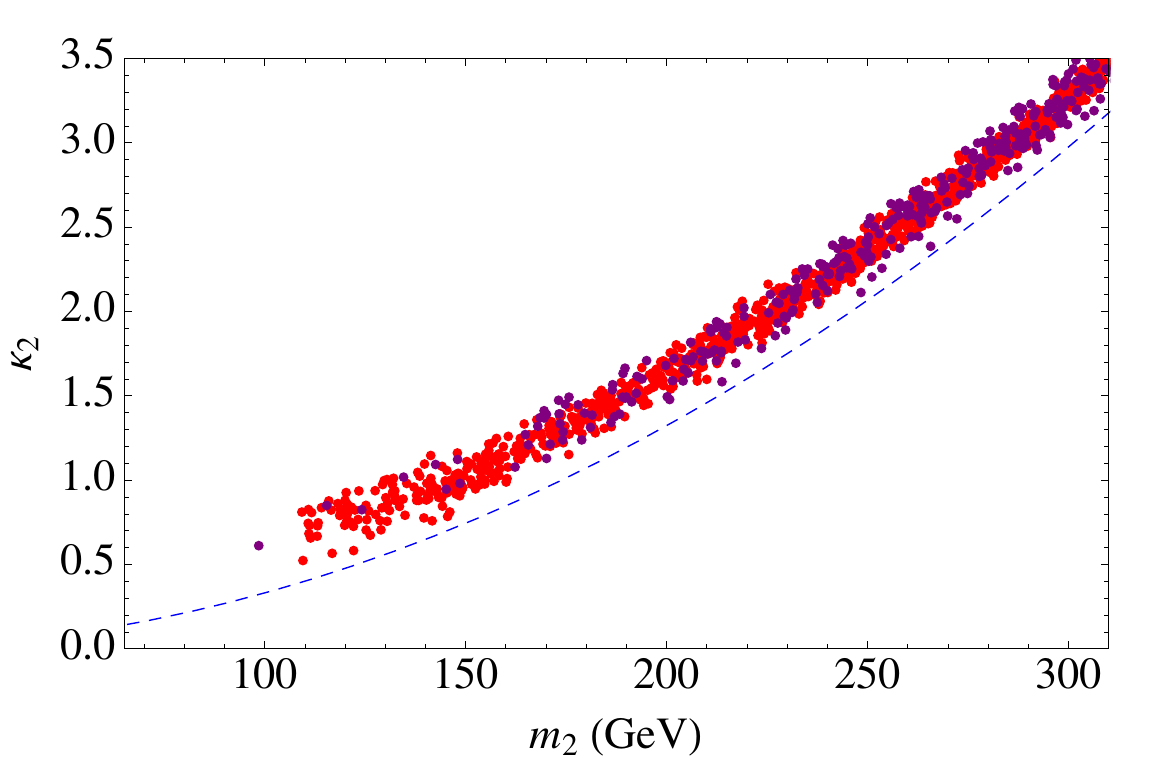}
\caption{Scatter plot depicting correlations between $m_2$, the physical mass of $S_2$, and $\kappa_2$, its coupling to $H_2$, for models that achieve a first-order phase transition.  The dashed blue curve denotes values of $\kappa_2$ above which $m^2_{2,S}$, the bare mass of $S_2$ must be tachyonic.
The allowed region does not depend sensitively on whether we employ the full thermal potential described in~\cite{Quiros:1999jp} (purple) or the leading order potential comprised of the zero temperature tree level potential plus thermal masses (red).
}\label{PhT}
\end{figure}

The presence of a vacuum at $\langle S_2 \rangle \neq 0$ requires that the thermal mass for $S_2$ be tachyonic during the EWPT.  
This may be possible if the zero temperature bare mass squared $m_{2, S}^2$ is sufficiently tachyonic; moreover, this instability will be enhanced if the operators $|S_1|^2|S_2|^2$ and $|H_1|^2|S_2|^2$ have negative coefficients, $\lambda_{3,S}$ and $\kappa_3$, respectively.  At very high temperatures, the temperature corrections to the thermal masses dominate over the zero temperature masses, so the signs of the quartic interaction terms dictate whether $\langle S_2 \rangle \neq 0$ at very high temperatures \cite{Weinberg:1974hy, Mohapatra:1979vr}. 
If these quartic interactions have negative couplings, then the conditions 
\bea
\lambda_{1,S} \lambda_{2,S} > \lambda_{3,S}^2 , \qquad
\lambda_{1,H} \lambda_{2,S} > \kappa_{3}^2,
\eea
must be satisfied in order for the potential to be bounded from below.  In general, the couplings are also bounded from above in magnitude by perturbativity.  For sufficiently large quartic couplings between the DM and the SM, running effects can push the model into a non-perturbative regime just above the weak scale \cite{Gonderinger:2012rd,EliasMiro:2012ay,Cheung:2012nb}.

To investigate the viable parameter space of our model we scan over all points consistent with a strongly first-order EWPT.  Our results are presented in \Fig{PhT}, employing the full potential described in \Eq{zeropot} and \Eq{thermalpot} (purple dots), as well as the lowest order potential only including the zero temperature tree-level potential and thermal masses (red dots).
The physical mass of $S_2$ after electroweak symmetry breaking is denoted by $m_2$, while $\kappa_2$ denotes the coupling of $H_2$ to $S_2$.  In \Fig{PhT} we have scanned all remaining dimensionless couplings within the range $[-2,2]$, and all bare masses squared from $[-(500 \textrm{ GeV})^2,(500 \textrm{ GeV})^2]$, keeping only those points consistent with the EWPT and our desired particle spectrum.

In principle, $m_{2}$ is unbounded from above, but according to \Fig{PhT}, larger values of the mass correlate with larger values of $\kappa_2$.  This is required in order to provide a sufficiently large potential barrier between the $\langle S_2 \rangle \neq 0$ and $ \langle H_2 \rangle  \neq 0$ vacua, as is clear from \Fig{PhT1}.  We see that restricting to perturbative values for $\kappa_2 \lesssim \sqrt{4\pi}$ implies a loose upper bound, $m_{2}\lesssim 300\,$GeV.  
In general, the correlation between $m_{2}$ and $\kappa_2$ implies a large coupling of $S_2$ to the Higgs boson which would be excluded by DM direct detection~\cite{Aprile:2012nq} were it the DM. However, this is not a problem because $S_2$ is unstable and decays to $S_1$.

\subsection{Asymmetry Generation}

A critical aspect of electroweak cogenesis is that $U(1)_{\rm DM}$ is an exact symmetry of the Lagrangian which is broken spontaneously in the early universe.  As a consequence,  particle asymmetries in $S_1$ and $S_2$ originate on the walls of  bubbles which have nucleated during the EWPT.  After the phase transition, $U(1)_{\rm DM}$ is restored, and a net DM asymmetry remains.

As depicted in \Fig{pic}, the nucleated bubbles interpolate between the VEVs described in \Eq{eq:VEVs}.  Because of CP violation in the 2HDM, the Higgs doublets acquire space-time varying CP phases, 
$\theta_1  = \arg \langle H_1 \rangle$ and $ \theta_2  = \arg \langle H_2 \rangle$, which provide the initial seeds for the baryon and DM asymmetries.  To simplify our analysis, we assume a thick-wall limit in which diffusion effects~\cite{Joyce:1994bk, Cohen:1994ss} can be justifiably neglected.
In this regime, the adiabatic variations in the CP phases provide an effective chemical potential for baryon number~\cite{Cohen:1991iu,Cohen:1993nk,Giudice:1993bb} which is manifest after applying an appropriate set of space-time dependent field redefinitions. 
In particular, $\theta_2$ enters into the top quark Yukawa coupling while $\theta_{21} = \theta_2 - \theta_1$ enters into the scalar interaction terms in the square brackets of \Eq{eq:VHS}, where  for simplicity we set $\epsilon_2=0$ throughout.  The effects of $\epsilon_2$ can of course be included in our analysis, but they will not qualitatively affect our final conclusions.  

To summarize, $\theta_2$ contributes an effective space-time modulating phase to the mass of the top quark, and likewise for $\theta_{21}$ and the $S_1$ and $S_2$ fields.
However, these phases can be removed by a field redefinition,
\bea
t &\rightarrow &e^{i \gamma_5 \theta_2 / 2}t\nonumber\\
S_1&\rightarrow& e^{i{\theta_{21}}/{2}} S_1 \nonumber\\
S_2&\rightarrow& e^{-i{\theta_{21}}/{2}} S_2 \ ,
\eea
at the cost of generating derivatively coupled terms, 
\bea
\Delta {\cal L}
= \frac{1}{2} \partial_\mu \theta_{2} (\bar t \gamma_5 \gamma^\mu t) + \frac{1}{2} \partial_\mu \theta_{21} (S_1^\dagger \overset{\leftrightarrow}{\partial^\mu} S_1-S_2^\dagger \overset{\leftrightarrow}{\partial^\mu} S_2).
\label{eq:chempot}
\eea
The terms involving time derivatives induce a potential difference which splits the energies of particles and anti-particles inside the bubble walls.   The induced chemical potentials for the top quark, $S_1$, and $S_2$ are
\bea
\mu_t = \dot \theta_2/2,\quad
\mu_{S_1}= -\mu_{S_2} = \dot \theta_{21}/2 \ ,
\eea
are crucial for generating the final asymmetries in baryons and DM, respectively.

The resulting baryon asymmetry arises as per usual in electroweak baryogenesis in the 2HDM.   
In the language of spontaneous baryogenesis, the chemical potential in \Eq{eq:chempot} together with top quark scattering in the wall produces a chiral charge asymmetry between left-handed and right-handed top quarks which is reprocessed by electroweak sphalerons to yield the final BAU.   
Due to the fast expansion of the bubble wall, this sphaleron reprocessing does not reach thermal equilibrium.
The comoving baryon number asymmetry evolves toward the thermal value dictated by chemical equilibrium according to the Boltzmann equation~\cite{Dine:1990fj}, 
\begin{eqnarray}\label{baryon} 
\frac{d \Delta_{\rm B}}{dt} \sim \Gamma_{\rm sph} \times  \mu_{t} \times T_c^2 \ ,
\label{eq:DB}
\end{eqnarray}
where $\Gamma_{\rm sph} \simeq 120 \alpha_w^5 T_c$ is the electroweak sphaleron rate.

Like the BAU, the DM asymmetry is directly connected to the chemical potentials induced by the space-time variation of the Higgs CP phases.  
There are two classes of microscopic processes that can generate asymmetries in the dark sector.
The first one is the scattering process $H_1^\dagger H_2 \to S_1 S_2^\dagger$, which generates equal and opposite asymmetries in $S_1$ and $S_2$ from the bubble wall, 
but no net $U(1)_{\rm DM}$ charge. The second class generate asymmetries in $S_1$ and $S_2$ independently, {\it e.g.}~via $t g \rightarrow t S_1, t S_2$, which can happen through mixing of $H_2$ with $S_1$ or $S_2$ on the wall.
Naively, the first class cannot contribute to a net $U(1)_{\rm DM}$ asymmetry, since all $S_2$ particles eventually decay to $S_1$.
However, as we show in the next subsection, the asymmetry in $S_2$ will be strongly washed out by after effects of the first-order phase transition,
while the asymmetry in $S_1$ will be preserved due to a small value for $\epsilon_1$.
Therefore, the final DM asymmetry is actually given by $\Delta_{\rm DM} = \Delta_{S_1}$.
We find the dominant such process of $S_1$ generation is the scattering $H_1^\dagger H_2 \to S_1 S_2^\dagger$, 
which occurs at a rate $\Gamma_{\rm scatt} \sim |\epsilon_1|^2 T_c/(4\pi)$.
In analogy with \Eq{baryon}, the comoving number density of DM is 
\begin{eqnarray}\label{dm}
\frac{d \Delta_{\rm DM}}{dt} =\frac{d \Delta_{S_1}}{dt}  \sim \Gamma_{\rm scatt} \times \mu_{S_1} \times   T_c^2 \ .
\label{eq:DDM}
\end{eqnarray}
There exist a number of subdominant processes which also produce an $S_1$ asymmetry.  For example, one can consider processes such as $H_1 Z \to H_1 S_1$ through a $S_1$ VEV insertion, which is induced on the bubble wall (of order $\epsilon_1$, see \Fig{pic}, however this is suppressed by the weak coupling constant. 
Another process is $t g \rightarrow t S_1$, which transfers the asymmetry from the top quark to $S_1$, via the $S_1$--$H_2$ mixing on the bubble wall.
The associated rate is suppressed by $|\epsilon_1|^2$, $\alpha_s$, and $1/\tan^2\beta$ from the $\langle H_1 \rangle$ VEV insertion.

The observed abundance of DM and baryons in the present day~\cite{Ade:2013zuv} implies that $\Omega_{\rm DM} / \Omega_{\rm B} \simeq 5.4$, so 
\begin{eqnarray}\label{adm}
|\Delta_{\rm DM} |= |\Delta_{S_1} |= | \Delta_{\rm B} |\left[\frac{5.4\,\rm GeV}{m_1}\right] \ .
\label{eq:coinc}
\end{eqnarray}
If we define the wall passage time, $\delta t_{\rm wall}$, to be the ratio of the width and velocity of the bubble wall, then clearly $\delta t_{\rm wall}\ll \Gamma_{\rm sph}^{-1}, \Gamma_{\rm scatt}^{-1}$.  \Eq{baryon} and \Eq{dm} at leading order in $\delta t_{\rm wall}$, together with \Eq{eq:coinc}, implies that
\begin{eqnarray}\label{adm2}
|\epsilon_1| \sim 2\times10^{-3} \left[ \frac{100\,\rm GeV}{m_{1}} \right]^{\frac{1}{2}} \left[ \frac{\Delta\theta_2}{\Delta \theta_{21}} \right]^{\frac{1}{2}} ,
\end{eqnarray}
in order to accommodate the observed BAU and ADM abundance.  Here $\Delta \theta_{2}$ and $\Delta {\theta_{21}}$ denote the changes in $\theta_2$ and $\theta_{21}$ across the bubble wall, which take on ${\cal O}(1)$ values if there is large CP violation in the scalar potential.   As noted in \Sec{sec:EWPT}, the Higgs CP phases scale as  $1/\tan \beta$ and become unphysical in the infinite $\tan \beta$ limit.
Here $\epsilon_1$ is required to be small by \Eq{adm2} because sphaleron processes are relatively weak and \Eq{baryon} and \Eq{dm} require $\Gamma_{\rm sph} \sim \Gamma_{\rm scatt}$.
Lastly, we note that \Eq{adm2} is wholly insensitive to the detailed phase transition temperature, bubble wall width and velocity---these quantities conveniently cancel between the ratio of baryon and DM asymmetries.

\subsection{Evolution of the Condensate}\label{subsec:condensate}

Like all mechanisms of particle asymmetry generation, electroweak cogenesis is subject to important washout effects.  Indeed, the restoration of $U(1)_{\rm DM}$ inside the nucleated bubble is not a perfect process.  Within bubbles of true vacua,
the $\langle S_2\rangle$ condensate will oscillate about zero before it eventually settles to the terminal vacuum at $\langle S_2 \rangle = 0$.
During this oscillatory phase, $S_2$ number is violated and any associated asymmetry will be washed out by particle oscillations, $S_{2}\leftrightarrow S_2^*$, which occur at a rate of
$\Gamma_{2, \rm osc} \sim \lambda_{2,S}^2 |\langle S_2\rangle|^2 /m_2$.   
Meanwhile, the amplitude of the condensate is damped by decay processes, $\langle S_2\rangle\to hh, WW, ZZ, \textrm{ or } b \bar b$, provided they are kinematically allowed. The corresponding decay rate of $S_2$ to the SM Higgs is $\Gamma_{2,\rm dec} \sim \kappa_2^2 |\langle S_2\rangle |^2/(8\pi m_2)$.
The scattering process $S_2 + h\to S_2+h$ has a rate $\Gamma_{2,\rm evap} \sim {\kappa_{2}^2 T}/{(8\pi)}$, and will decohere and eventually evaporate the $\langle S_2 \rangle$ condensate into particles.    

In principle, the asymmetry in $S_1$ particles will undergo similar washout effects from the oscillation, damping, and evaporation of the $\langle S_1 \rangle$ condensate.
However, the oscillation and decay effects are strongly suppressed by $\epsilon_1$, which according to \Eq{adm2} must be small to accommodate the observed baryon and DM asymmetries.  For example, $\Gamma_{1, \rm osc} \sim \lambda_{1,S}^2 |\langle S_1\rangle|^2/m_1$ is tiny because $\langle S_1\rangle$ is induced sub-dominantly from $\langle S_2\rangle$ proportional to $\epsilon_1$; $\Gamma_{1, \rm dec}$ is similarly suppressed.
On the other hand, the evaporation of $\langle S_1 \rangle$ by scattering with the Higgs particles is unrelated to $\epsilon_1$ and thus $\Gamma_{1,\rm evap}$ is sizable.
Therefore, it is straightforward to arrange the following hierarchy among the relevant rates,
\begin{eqnarray}\label{rateorder}
&\Gamma_{2,\rm osc} > \Gamma_{2,\rm evap},  \Gamma_{2,\rm dec}  \gg \rm Hubble  & \nonumber \\
&\Gamma_{1, \rm evap} \gg \rm Hubble \gg \Gamma_{1, \rm osc}, \Gamma_{1, \rm dec} \ , &
\end{eqnarray}
so the $S_2$ asymmetry is efficiently erased but the $S_1$ asymmetry is preserved.  The reason for the hierarchy in $S_2$ rates is that condensate decay and evaporation are effectively higher body processes than condensate oscillation.  

In the above discussion we have neglected the terms in \Eq{eq:VS} with the coefficients $\lambda_{4,S}$ and $m^2_{3,S}$.  While these operators preserve $U(1)_{\rm DM}$, they explicitly break the orthogonal ``axial'' symmetry that acts oppositely on $S_1$ and $S_2$.
If these interaction terms are large, then the washout of $S_2$, together with the induced mixing between $S_1$ and $S_2$, will wash out $S_1$.
Requiring that the effects of these interactions are less than that of $\epsilon_1$, we demand that  $|\lambda_{4,S}| <|\epsilon_1|$ and $|m^2_{3,S}/(m^2_2-m^2_1)|<|\epsilon_1|$.

Another potential contribution to the DM asymmetry is from the complex condensate itself.  
As is well-known, such asymmetries naturally arise in models with dynamical scalar fields, {\it e.g.}~in Affleck-Dine baryogenesis or cogenesis~\cite{Affleck:1984fy, Dine:1995kz, Enqvist:2003gh,Cheung:2011if}.  Indeed, our condensate carries a time dependent phase, corresponding to ``spinning'' of the $S_1$ and $S_2$ around the origin of field space.  As a consequence, these field configurations carry intrinsic particle asymmetries given by
\bea
n_{S_1} = |\langle S_1\rangle|^2 \dot \varphi_1, \qquad n_{S_2} = |\langle S_2\rangle|^2 \dot \varphi_2,
\eea
where $\varphi_{1}=\arg\langle S_1\rangle$ and $\varphi_{2}=\arg\langle S_2\rangle$.
The condensate asymmetry finally evaporates into the corresponding particle asymmetry, and experience similar washout effects as above.
We observe that if $\lambda_{4,S}$ and $m^2_{3,S}$ are negligibly small, then the condensate asymmetries in $S_1$ and $S_2$ are induced by the $\epsilon_1$ term.  Due to conservation of $U(1)_{\rm DM}$, the equations of motion require that $\partial_t (|\langle S_1\rangle|^2 \dot \varphi_{1}) = - \partial_t (|\langle S_2\rangle|^2 \dot \varphi_{2})$.
Therefore, the asymmetries stored in the $S_1$ and $S_2$ condensates are equal and opposite, 
of order $|\epsilon_1|^2/\tan^2\beta$, and inversely proportional to $\delta t_{\rm wall}$. 
Comparing with \Eq{dm}, we find that the asymmetry contained directly within the condensate is subdominant to the asymmetry induced from particle scattering provided the bubble wall is relatively slow and $\tan\beta$ is large.

\section{Experimental Constraints}\label{sec:exp}

\subsection{Direct Detection}

ADM requires strong annihilation channels to deplete the symmetric abundance of DM during thermal freeze-out.  In particular, the thermally averaged DM annihilation cross-section is bounded by
\bea
\langle \sigma v\rangle > 3\times 10^{-26}\textrm{ cm}^3/\textrm{s}.
\label{eq:subthermal}
\eea
To accommodate \Eq{eq:subthermal}, conventional models of ADM typically introduce new light degrees of freedom into which DM can annihilate.  In principle, these light states can induce experimentally observable signals in DM-nucleon scattering, but this connection is highly model dependent, so ADM lacks a universal prediction for direct detection.  In contrast, electroweak cogenesis utilizes weak scale ADM which couples to the SM through the Higgs portal and avoids light mediators altogether.  Thus, the very same couplings which must be sufficiently large to satisfy~\Eq{eq:subthermal} are also subject to stringent direct detection limits.  

For simplicity, consider the limit of our model in which the heavier Higgs doublet is decoupled, yielding the SM plus two complex singlets~\cite{Barger:2008jx}.  Without resorting to higher dimension operators, this theory does not possess the requisite CP violation to accommodate baryogenesis.  Even neglecting this issue, the parameter space consistent with~\Eq{eq:subthermal} and XENON100 is quite narrow~\cite{Djouadi:2011aa}.  In the allowed region, the annihilation cross-section of DM cannot be much stronger than the thermal case, so the relic DM carries a sizable symmetric abundance---the DM is not fully asymmetric.

Next, we consider the regime in which both Higgs doublets have weak scale masses.  
Because  CP is explicitly broken in the Higgs sector, the CP even and odd scalars will mix among each other.  Hence, the neutral gauge eigenstates, $(H_1^0, H_2^0, A)$, are related to the mass eigenstates, $(h_1,h_2,h_3)$, by a general orthogonal matrix $R_{ij}$~\cite{Accomando:2006ga,WahabElKaffas:2007xd}.
It has been shown~\cite{Shu:2013uua} recently that CP violation for the electroweak baryogenesis window is still consistent with the electric dipole moment constraints, as well as the measurement of Higgs properties at the LHC.
In the presence of multiple Higgs states~\cite{Bai:2012nv} it is possible to arrange for destructive interference among the contributing Feynman diagrams in DM direct detection.  Critically, in these parameter regions the DM annihilation channels are not suppressed, so \Eq{eq:subthermal} can be accommodated.

The effective Lagrangian describing the couplings of DM to quarks is obtained by integrating out the Higgs bosons,
\begin{eqnarray}\label{LDMQ}
\mathcal{L}_{\rm eff} &=& \frac{m_u}{\Lambda_u^2} S^\dagger S \bar u u + \frac{m_d}{\Lambda_d^2} S^\dagger S \bar d d  \\
\frac{1}{\Lambda_u^2} &=& \sum_{i=1}^3 \frac{\lambda_{i}R_{i2}}{m_{h_i}^2 \sin\beta}  \ , \ \ 
\frac{1}{\Lambda_d^2} = \sum_{i=1}^3 \frac{\lambda_{i}R_{i1}}{m_{h_i}^2 \cos\beta} \ ,
\end{eqnarray}
where $\lambda_{i} = \kappa_{1} \cos\beta R_{i1} + \kappa_{4} \sin\beta R_{i2}$.
The cross-section for spin-independent DM-nucleon scattering is 
\begin{eqnarray}
\sigma_{\rm SI} = \frac{\mu_N^2 m_N^2}{4\pi m_1^2} \left( \sum_q {f_{T_q}^{(N)}}/{\Lambda_q^2} \right)^2 ,
\end{eqnarray}
where $m_N$ is the nucleon mass and $\mu_N$ is the reduced mass of the DM-nucleon system. 
For nucleon matrix elements $\langle N| m_q \bar q q |N\rangle = m_N f_{T_q}^{(N)}$,
we use the latest lattice results~\cite{Giedt:2009mr}.

\begin{figure}[t]
\centerline{\includegraphics[width=1.0\columnwidth]{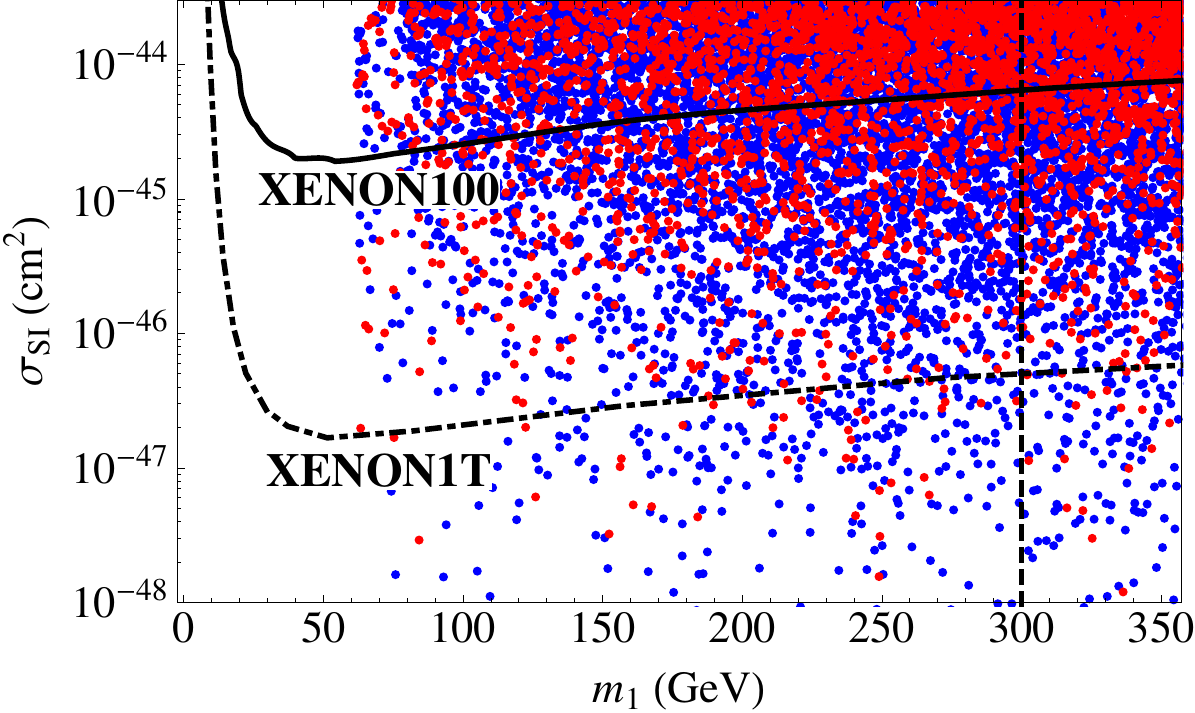}}
\caption{Scatter plot depicting model points consistent with the observed relic abundance of DM, together with present and future limits from XENON100 (solid black) and XENON1T (dashed-dotted black), respectively~\cite{Aprile:2012zx}.  CP conserving (red) and violating (blue) models are shown, along with the rough upper bound on the DM mass arising from the requirement of a strong first-order EWPT (dashed black). }\label{cp}
\end{figure}

\Fig{cp} depicts parameter points which satisfy the relic density constraints~\cite{Ade:2013zuv} together with direct detection limits from XENON100~\cite{Aprile:2012nq}.
To compute the thermal symmetric DM relic density, we have taken into account the dominant annihilation channels into $WW, ZZ, \textrm{ and } b\bar b$ including the three-body annihilation channels.  Due to the limits on the invisible Higgs width, $m_1 \gtrsim m_h/2$, while $m_1$ is bounded from above because $m_1 < m_2$ and $m_2$ is limited by the phase transition conditions in \Fig{PhT}. It is clear that CP violation in the Higgs sector allows for more natural cancellations in the direct detection cross-section,
and opens richer parameter space for our scenario.

\subsection{Collider Signatures}

We now discuss the collider phenomenology of our model.   It will be quite difficult to directly observe the complex scalars in our model, since they only couple to the SM through the Higgs doublets.  As such, our best prospects for experimental signs at the LHC rely on the 2HDM sector of the theory.  ATLAS~\cite{heavyH} has placed substantial constraints on the neutral component of the heavy Higgs decaying to $W^+W^-$.  These limits depend on $\tan \beta$ and $\alpha$, which denotes the mixing angle between the $H_1$ and $H_2$ mass eigenstates.
For generic values of $\alpha, \beta$, the lower limit on the mass is 200--300\,GeV, except near the decouplng limit, $\alpha=\beta-\pi/2$, when the heavy Higgs does not couple to electroweak gauge bosons.

In the presence of CP violation in the Higgs sector, all of the neutral scalars $h_{1,2,3}$ can decay into a pair of gauge bosons.
It is also crucial to measure the CP violating effects in the Higgs productions and decays.
The prospects of other decay channels have been recently studied in~\cite{Craig:2013hca}, as well as possible modifications of Higgs properties from additional singlets~\cite{Cheung:2013bn}.
The charged scalar may also be probed via associated production with gauge bosons~\cite{Basso:2013wna}, and is also constrained by the flavor violation process $b\to s\gamma$ to be heavier than 295\,GeV~\cite{Misiak:2006zs} in the case of Type-II 2HDM.

The singlet scalars in our model are difficult to see because they only couple via the Higgs portal.  Because their masses lie above half the Higgs boson mass, they cannot be probed through invisible width measurements.  On the other hand, the production of heavy Higgses followed by their subsequent decays will lead to missing energy events, like $pp\rightarrow h^+ h^- \rightarrow W^+ W^- h_2 h_2$ followed by the invisible decay of $h_2$ into a DM pair.
In this case, the collider signatures will be similar to that of the inert Higgs doublet model~\cite{Dolle:2009ft}.  
Therefore, through the electroweak production of heavier Higgs states it may be possible to probe the dark sector of this model in future runs of the LHC.

\section{Future Directions}\label{sec:conclusions}

Electroweak cogenesis is a simple framework for baryogenesis and asymmetric dark matter generation at the EWPT.    Its key ingredient is an exact $U(1)_{\rm DM}$ symmetry that is spontaneously broken in the early universe by the VEV of a $U(1)_{\rm DM}$ charged field.  After the EWPT, $U(1)_{\rm  DM}$ is restored, together with $U(1)_{\rm B}$.  As the result of $U(1)_{\rm B}$, $U(1)_{\rm DM}$, and CP violating interactions in the vicinity of the nucleated bubble walls, baryon and DM asymmetries are simultaneously generated.   Since the ratio of $n_{\rm B}$ and $n_{\rm DM}$ is controlled by the parameters of the theory, the DM mass can be at the weak scale rather than the GeV scale, in contrast with conventional models of ADM.  As a consequence, Higgs portal interactions are sufficient to deplete the symmetric component of our DM particle, and light mediators need not be introduced.   Electroweak cogenesis offers a minimal realization of ADM using just the 2HDM augmented by two complex singlet scalars.  The present analysis leaves open a number of interesting possibilities for future work, which we now discuss.  

A full appraisal of electroweak cogenesis calls for a more rigorous study of the phase transition and the process of asymmetry generation.
For example, to quantitatively evaluate the strength of the phase transition, it would be useful to analyze the complete field dynamics including the effects of back-reaction from $H_1$ and $S_1$ on the thermal potential.   A more quantitative analysis would also entail a higher order calculation of a gauge-invariant effective thermal potential~\cite{Patel:2011th} or a non-perturbative evaluation of the phase transition on the lattice.
Our determination of the CP violating sources could also be refined as in~\cite{Riotto:1998zb,Lee:2004we}, along with the evolution of the number density asymmetries via particle diffusion~\cite{Huet:1995sh,Lee:2004we} during the EWPT.

Another subject which warrants further study connects with the role of the scalar condensate in electroweak cogenesis.  Thus far, we have focused on a scenario in which asymmetry generation arises dominantly from particle scattering in the background of the nucleated bubble walls.  However, as noted in \Sec{subsec:condensate}, there is another source of DM asymmetry arising from the coherent scalar field configurations which comprise the walls themselves.  In principle, the condensate can contribute an intrinsic and dominant source of the DM asymmetry, as is the case in other models of asymmetry generation \cite{Cheung:2012im, Affleck:1984fy, Dine:1995kz, Enqvist:2003gh}.

Lastly, it would be interesting to consider the viability of electroweak cogenesis within more general theories beyond the SM.  Indeed, the model presented in this paper relies on fundamental scalars, and as a result carries the usual burden of radiative instability.  It is therefore an interesting question whether electroweak cogenesis can be achieved within the context of technically natural UV completions such as supersymmetry or composite Higgs theories.  Conveniently, many variations of these models predict additional singlet scalars at the weak scale which could furnish a viable ADM candidate.

\begin{center} {\it Acknowledgements}\end{center}
 
We thank Pavel Fileviez-Perez, Xiangdong Ji, Rabi Mohapatra, Michael Ramsey-Musolf, Shmuel Nussinov, Goran Senjanovic, Jing Shu, Mark Wise, and Kathryn Zurek for valuable comments and discussions.
This work is supported by the Gordon and Betty Moore Foundation through Grant \#776 to the Caltech Moore Center for Theoretical Cosmology and Physics, and by the DOE Grant DE-FG02-92ER40701.

\end{document}